\documentclass{article}
\usepackage{emulateapj}

\def\msun{{\rm M_{\odot}}}
\def\rsun{{\rm R_{\odot}}}

\def\be{\begin{equation}}
\def\ee{\end{equation}}
\begin{document}

\title{ULTRALUMINOUS X--RAY SOURCES IN EXTERNAL GALAXIES}

\author{ A.R.~King\altaffilmark{1}, M.B.~Davies\altaffilmark{1},
M.J.~Ward\altaffilmark{1}, G. Fabbiano\altaffilmark{2} \&
M. Elvis\altaffilmark{2}}

\altaffiltext{1} {Department of Physics and Astronomy, University of
Leicester, Leicester LE1 7RH, U.K.; ark@star.le.ac.uk}
\altaffiltext{2} {Harvard--Smithsonian Center for Astrophysics, 60
Garden Street, Cambridge, MA 02138}

\begin{abstract}

We investigate models for the class of ultraluminous non--nuclear
X--ray sources (ULXs) seen in a number of galaxies and probably
associated with star--forming regions. Models where the X--ray
emission is assumed to be isotropic run into several difficulties. In
particular formation of 
sufficient numbers of 
the required ultramassive black--hole X--ray
binaries is problematic, and the likely transient behaviour of the
resulting systems is not in good accord with observation. The
assumption of mild X--ray beaming suggests instead that ULXs may
represent a shortlived but extremely common stage in the evolution of
a wide class of X--ray binaries. The best candidate for this is the
phase of thermal--timescale mass transfer inevitable in many
intermediate and high--mass X--ray binaries. This in turn suggests a
link with the Galactic microquasars.  The short lifetimes of
high--mass X--ray binaries would explain the association of ULXs with
episodes of star formation. These considerations still allow the
possibility that {\it individual} ULXs may contain extremely massive
black holes.

\end{abstract}

\keywords{accretion, accretion discs
--- binaries: close  --- X-rays: stars}

\section{INTRODUCTION}

The existence in spiral galaxies of off--nuclear X--ray sources whose
luminosities appear significantly larger than the Eddington limit for
a $1\msun$ object has been known for some time (Fabbiano, 1989).
These sources are distinct from the weak AGN present in many spiral
galaxies known as LINERs (Ho et al. 1997), although in at least one
case (M33, cf Dubus et al, 1997) they may be confused with
AGN. Recently considerable effort has been devoted to interpreting the
properties of these `ULXs' (= ultra--luminous compact X--ray sources,
e.g Makishima et al, 2000 and references therein). A key to
understanding their nature may be that they appear to occur
preferentially, although not exclusively, in regions of star formation
(Zezas, Georgantopoulos \& Ward 1999; Roberts and Warwick 2000,
Fabbiano, Zezas \& Murray, 2001). 
In this paper we investigate models
for the ULXs.

Bright, non--nuclear X--ray point sources in galaxies divide into two
groups: accreting neutron stars and black holes, and young supernova
remnants. The luminosities of the first group, but not the second,
are constrained by the Eddington limit:
\begin{equation}
L_X \la L_{\rm Edd} \simeq {4\pi GM_1m_pc\over \sigma_T}
\simeq 1.3\times 10^{38}m_1~{\rm erg\ s}^{-1},
\label{edd}
\end{equation}
where $\sigma_T$ is the Thomson cross--section and $m_1$ is the
accretor mass $M_1$ in $\msun$. This constraint applies to {\it any}
non--explosive source, whether powered by accretion or other means
such as nuclear burning.

Evasions of the limit are possible, but rare. In the transient system
A0538--66 a magnetic neutron star accretes from the wind of a Be--star
companion. The system sometimes has super--Eddington luminosities
$L_X \simeq 10^{39}~{\rm erg\ s}^{-1}$ (White \& Carpenter, 1978)
but these may result from the reduction in the electron
scattering cross--section below $\sigma_T$ in the $\sim 10^{11}$~G
magnetic field of the neutron star pervading the accretion
columns. 
%For accreting sources respecting the limit, with $m_1 \simeq
%1.4$ for neutron stars, and $m_1 \la 10$ for typical black--hole soft
%X--ray transients (e.g. Charles, 1999) one generally expects $L_X \la
%1\times 10^{39}~{\rm erg\ s}^{-1}$.
%
Thus if we regard the observed variability of ULXs as ruling out the
identification as supernova remnants, a straightforward interpretation
as non--explosive sources requires black holes with masses $M_1 \ga 50
- 100\msun$, accreting at rates which on occasion can exceed $\sim
10^{-6}\msun~{\rm yr}^{-1}$.
As we shall see, while individual ULXs could harbour such masses,
there are major difficulties with such a picture as an explanation for
the ULXs as a class. Accordingly we consider models in which the
X--ray emission is assumed to be significantly beamed. In this case
%
%
%There are two major difficulties to overcome here. 
%Descent from very
%massive stars is problematic because of the prodigious mass loss such
%stars probably suffer; it is often claimed that this would reduce the
%masses of any black hole remnant to $\la 10\msun$. The second problem
%is the mass supply required to power the observed luminosities.
%
%We shall examine these problems in what follows. However in view of
%the potential difficulties, we also discuss a completely different
%line of explanation for ULXs. 
%The very high inferred black hole masses
%come directly from the assumption that the observed flux density is
%part of an isotropic emission pattern. If instead the emission is
%beamed into a restricted solid angle the resulting mass limits are
%reduced. Of course this comes at the price of requiring more sources,
%albeit of lower luminosity. We arrive at two possible types of model
%for ULXs. If they are unbeamed, the most likely picture is an
%extremely massive X--ray binary, in which a $\sim 100\msun$ black hole
%accretes from a stellar companion via a massive disc; accretion
%through the disc is episodic because of the thermal--viscous
%instability, and the ULX phase corresponds to (very long) outburst
%phases. If ULXs are significantly beamed, this state 
%
%
ULXs
may correspond to a relatively shortlived 
but common
epoch of the evolution of close
intermediate-- or high--mass X--ray binaries, perhaps 
the
thermal--timescale mass transfer 
phase following the normal X--ray phase.
%
%
%on to a moderate--mass black hole or neutron star from a more
%massive companion.

\section{LUMINOSITIES, LIFETIMES, MASSES, BIRTHRATES}

We first consider the restrictions placed by observation on accretion
models for the ULXs. We assume that a compact object of mass $M_1$
accretes from a mass reservoir (e.g. a companion star) of mass $M_2$. 
We denote the mean observed number of ULXs per galaxy as $n$, the
beaming factor as $b$ ($ = \Omega/4\pi$, where $\Omega$ is the solid
angle of emission), the duty cycle (= time that the source is active
as a fraction of its lifetime) as $d$, and define an `acceptance rate' $a$
as the ratio of mass accreted by $M_1$ to that lost by $M_2$, i.e. the
mean accretion rate $\dot M_1 = a(-\dot M_2)$. We further define $L_{\rm
sph}$ as the apparent X--ray (assumed bolometric) luminosity of a
source, given by the assumption of isotropic emission, and let $L_{40}
= L_{\rm sph}/10^{40}\ {\rm erg\ s}^{-1}$. From these definitions it
follows that the luminosity 
\begin{equation}
L = bL_{\rm sph} = 10^{40}bL_{40}\ {\rm erg\ s}^{-1}
\label{l}
\end{equation}
and the minimum accretor mass if the source is not to exceed the Eddington
limit is
\begin{equation}
M_1 \ga 10^2bL_{40}\msun.
\label{m}
\end{equation}
The total number of such sources per galaxy is
\begin{equation}
N = {n\over bd}
\label{n}
\end{equation}
with a minimum mean accretion rate during active phases of
\begin{equation}
\dot M_{\rm active} = {\dot M_1\over d} = -{\dot M_2a\over d} >
10^{-6}bL_{40}\ \msun\ {\rm yr}^{-1}. 
\label{mdoto}
\end{equation}
The mass loss rate from $M_2$ is thus
\begin{equation}
-\dot M_2 >
10^{-6}{bd\over a}L_{40}\ \msun\ {\rm yr}^{-1},
\label{mdot}
\end{equation}
and the lifetime of a source is
\begin{equation}
\tau = -{M_2\over\dot M_2} \la
10^{6}{m_2a\over bdL_{40}}\ {\rm yr},
\label{tau}
\end{equation}
with $m_2 = M_2/\msun$, leading to a required birthrate per galaxy 
\begin{equation}
B = {N\over\tau} \ga {n\over bd}.{bdL_{40}\over 10^6m_2a} = 
10^{-6}{nL_{40}\over m_2a}\ {\rm yr}^{-1}.
\label{b}
\end{equation}
%MBD
%The important point to note here is that the required birthrate is
It is important to note here that the required birthrate is
independent of beaming (and duty cycle): the greater intrinsic source
population $N$ required by $bd < 1$ (cf eq. \ref{n}) is compensated by
their longer lifetimes (cf eq. \ref{tau}).

\section{UNBEAMED MODELS}

For an unbeamed model we set $b=1$, and recover from (\ref{m}) the
requirement $M_1 \ga 10^2L_{40}$. 
%we ignore for the moment the
%possible difficulty of forming such an object, and investigate if a
%viable model for ULXs can be constructed assuming the existence of
%black holes of the required masses. 
We note immediately that some
ingenuity is required (cf Makishima et al., 2000) to make these masses
compatible with the characteristic observed X--ray
temperatures 1 -- 2 keV, whereas these are natural values for the
lower masses we shall find below for beamed models.
%
%For unbeamed models, 
With $b=1$ 
%equation 
(\ref{mdoto}) gives $\dot M_{\rm active}
\ga 10^{-6}L_{40}\ \msun\ {\rm yr}^{-1}$. This effectively forces the
mass reservoir to be a companion star; 
except for extremely high black hole masses $M_1 \ga 3\times 10^4\msun$,
Bondi accretion from even a
relatively dense part of the interstellar medium is inadequate, giving
a rate
\begin{equation}
\dot M_{\rm Bondi} \simeq 1\times 10^{-11}{m_{100}^2(\rho/10^{-24}\ 
{\rm g\ cm}^{-3})\over u_{10}^2 + c_{10}^2}\ \msun\ {\rm yr}^{-1}
\label{bondi}
\end{equation}
where $m_{100} = M_1/100\msun, \rho$ is the mass density of the
ambient interstellar medium, and $u_{10}, c_{10}$ are the relative
speed of the hole and the local ISM, and the local ISM sound speed,
both in units of 10~km s$^{-1}$. 
While individual ULXs might contain
black holes of masses $M_1 \ga 3\times 10^4\msun$, it seems improbable
that a galaxy like the Antennae should contain about 10 accreting examples.
Unbeamed models for the ULX class thus have to
invoke a 
class of
extremely massive X--ray binaries. As we shall see, this may
be a potential problem. Accepting that suitable binaries could in
principle form, there are the usual constraints familiar from LMXB
evolution (cf Kalogera \& Webbink, 1996). Particularly important are
(a) the binary must be wide enough that the progenitor of the compact
star (here a $\sim 100\msun$ black hole) must be able to fit inside
its Roche lobe (otherwise it will provoke common--envelope evolution),
and (b) the binary must be able to provide the inferred minimum
accretion rate $\dot M_1 \sim 10^{-6}\msun\ {\rm yr}^{-1}$. Constraint
(a) immediately sets a scale, as main--sequence stars of masses $\ga
100\msun$ have radii $\ga 10^3\rsun$ (e.g. Stothers \& Chin, 1999). Using
Kepler's law, and assuming $M_1 >> M_2$, this implies binary periods
\begin{equation}
P \ga 1m_{100}^{-0.5}~{\rm yr}.
\label{p}
\end{equation}
We can compare this with the critical period beyond which the
accretion disc around the black hole cannot be thermally stable, and
the system must therefore be transient. From King (2000) we find
\begin{equation}
P_{\rm crit} \sim 4m_{100}^{1/8}m_2^{1/8}~{\rm d}.
\label{pcrit}
\end{equation}
We see that unbeamed ULXs must be transient. Hence the inferred $\dot
M_1 \sim 10^{-6}\msun\ {\rm yr}^{-1}$ now refers to the outburst state
only: this is advantageous, as persistent mass transfer rates $-\dot
M_2$ of this order would have been difficult to explain. To fill the
Roche lobe in a binary with period given by (\ref{p}) requires an
extended star (note that the binary period essentially fixes the mean
density of this star uniquely, cf e.g. Frank et al., 1992, Ch. 4). From
the formulae of King (1988) (cf Ritter, 1999) we see that an evolved
star with helium core mass $M_c \sim 0.4 - 0.5\msun$ will fill the
Roche lobe, independently of the total donor mass $M_2 > M_c$. Mass
transfer is driven by the nuclear expansion of the star, at the rate
\begin{equation}
-\dot M_2 \simeq 1\times 10^{-7}P_{\rm yr}^{0.93}m_2^{1.47}\msun~{\rm yr}^{-1}
\label{trans}
\end{equation}
where $P_{\rm yr}$ is the binary period in years (cf King, Kolb \&
Burderi, 1996, eq 7). Thus even a duty cycle $d$ as long as 10\% would
yield mean accretion rates of the required order, i.e. $\dot M_1 =
\dot M_{\rm active} = \dot M_1d^{-1} \sim 10^{-6}\msun\ {\rm yr}^{-1}$
during outbursts.

%The nature of the predicted outbursts is crucial to successful
%modelling of the ULX population. 
%While there are indications of
%variability, there is little apparent change in the ULX population
%between observations by ROSAT and {\it Chandra}. Accordingly the
%transient outbursts must last several decades. 
%The duration of the
%outburst is set by the viscous timescale of the maximal region of the
%disc irradiated by the central X--rays (King \& Ritter, 1998; Dubus et
%al., 1999). The estimates of King \& Ritter (1998) show that this
%region has a typical size
%\begin{equation}
%R_h \simeq 5\times 10^{12}(\dot M_{\rm active}/10^{-6}\msun\ {\rm
%yr}^{-1})^{1/2}~{\rm cm}.
%\label{rh}
%\end{equation}
%This is far smaller than the binary separation $\ga 10^3\rsun =
%7\times 10^{13}$~cm, so most of the accretion disc is not irradiated;
%this is a familiar feature of long--period transients. The associated
%viscous timescale $R_h^2/3\nu$ ($\nu = $ kinematic viscosity) is
%indeed 
%of the order of $10 - 20$~yr.
%so the predicted total outburst duration appears reasonable. 
However, 
applying
the simple irradiated--disc theory of King \& Ritter (1998) 
%
%
%(see also Dubus et al., 1999)
%
%
predicts that the accretion rate $\dot
M_{\rm active}$ should decay linearly from its initial peak down to
zero on 
%this 
%
%
a 10 --20~yr
timescale.
%(a linear decay of exactly this type is seen in
%GRO~J1744-28, the transient with the longest observed orbital period
%(12~d) (Giles et al., 1996). 
This is not easily compatible with
comparison of ROSAT and {\it Chandra} data. This may not be a crucial
objection to this type of unbeamed model for ULXs; the theory of
outbursts in large irradiated discs is complicated, even without
adding further difficulties such as radiation--induced disc warping.

%MBD
%However 
A more serious objection to unbeamed models is the one touched
on above, namely that they require a $\sim 100\msun$ black hole to
coexist in a binary with an evolved companion star. Mass loss from
very massive stars with non--zero metallicity is usually thought to be
so strong that the final black hole mass is much smaller than the
initial stellar mass (see Baraffe, Heger \& Woosley 2001 for a recent
view, and Papaloizou 1973 for a possible objection). Assuming coeval
formation of the two stars in the binary rules out a primordial origin
for the black hole progenitor, and would thus require a progenitor
with a mass $\gg 100\msun$. If the IMF within the stellar clusters is
close to that of Salpeter, we conclude that the number of stars formed
having masses $\geq 100\msun$ is a factor $\sim 100$ lower than the
number of stars having initial masses $10 \leq m \leq 100 \msun$. We
would therefore expect X--ray binaries containing neutron stars and
lower--mass black holes to outnumber markedly those containing
higher--mass black holes. The X--ray luminosity of the systems
observed in the Antennae (Fabbiano, Zezas \& Murray, 2001) contradicts
this, assuming that all systems are unbeamed, although the number of
low luminosity systems is not currently well known.

Alternatively, a $\sim 100\msun$ black hole may only recently have
gained a new stellar partner.  Such black holes may be produced within
dense clusters through the merger of lower--mass black holes (see for
example Lee 1993, 1995). Indeed this has been proposed as the origin
of the moderate--mass black hole inferred to be present in the central
regions of M82 (Matsushita et al 2000, Matsumoto et al. 2001, Kaaret
et al., 2001). Although it is possible that a moderate--mass black
hole produced in a central cluster of M82 has gained a stellar
companion by some dynamical process (tidal capture or via an exchange
encounter involving a binary, this scenerio is unlikely to work for
the systems observed in the Antennae where the ULXs are observed to be
strongly associated with the young star clusters which are located
some distance from the galactic nuclei (Fabbiano, Zezas \& Murray,
2001). Any massive BHs would therefore have to be formed within these
stellar clusters and not the nuclear clusters. To produce
moderate--mass black holes within a cluster via the successive merger
of lower--mass objects, the potential well of the cluster has to be
sufficiently deep to retain the black holes. This can be the case for
the stellar cluster in the nucleus of a galaxy but is not true for
globular clusters where the typical escape speed is far too low to
retain black hole binaries as they are hardened via encounters. This
has been suggested as the reason for the absence of black hole binary
systems in globular clusters (Sigurdsson \& Hernquist 1993, Kulkarni
et al 1993).

A population of $\sim 100\msun$ black holes originating from a much
earlier generation of effectively zero--metallicity stars seems
unlikely to explain the ULXs observed in the Antennae (Fabbiano, Zezas
\& Murray, 2001) as these black holes would be distributed throughout
the galactic halo and the probability of picking up stars from the
young stellar clusters via dynamical encounters within the last $\sim
10^7$ years is extremely low.

\section{BEAMED MODELS}

Since unbeamed models run into difficulties because of the required
black hole mass $\sim 100\msun$ and the need for a companion, we
consider the effect of assuming that the observed X--rays are mildly
beamed. The simplest candidate mechanism is the idea that the
accretion disc around an accreting black hole has much lower
scattering optical depth over a restricted range of solid angles
(e.g. the rotational poles) than in other directions.  Almost all the
emitted X--rays would therefore emerge in these directions. A beaming
factor $b \la 0.1$ would bring the required minimum accretor mass (cf
eqn \ref{m}) into the range $M_1 \la 10\msun$ commonly found in
dynamical measurements of X--ray binaries, particularly quiescent soft
X--ray transients (e.g. Charles, 1998), while $b \la 0.01$ would bring
$M_1$ down to neutron--star values. In addition this kind of disc
geometry, i.e. a thick disc with a central funnel, may actually
radiate a total luminosity in excess of the Eddington limit
(Jaroszynski, Abramowicz \& Paczynski, 1980; Abramowicz, Calvani \&
Nobili, 1980). Thus such modest $b$--values may allow quite large
apparent luminosities for perfectly standard black--hole or
neutron--star masses. The obvious implication is that beamed ULXs
might represent some short--lived phase in the evolution of a large
class of X--ray binaries: from eqn \ref{tau} we find $\tau \la
10^7m_2a/(b/0.1)d$~yr.

A good candidate for such a phase is an episode of thermal--timescale
mass transfer. These are extremely common, occurring when the donor
has a radiative envelope and is either (a) somewhat more massive than
the accretor, and/or (b) first fills its Roche lobe as it expands
across the Hertzsprung gap. In general both cases give rise to highly
super--Eddington mass transfer rates. Case (a) is unavoidable for
example in any neutron--star binary with an intermediate--mass ($\sim
2 - 4\msun$ donor): King \& Ritter (1999) and Podsiadlowski \&
Rappaport (2000) show that Cygnus X--2 is a survivor of such an
episode, in which $-\dot M_2$ reached values of order $\sim
10^{-6}\msun~{\rm yr}^{-1}$ and the excess mass transfer is simply
blown away from the system rather than resulting in common--envelope
evolution (see also King \& Begelman 1999, and Kolb et al., 2000).
Case (b) requires only a reasonably wide binary separation after
formation of the compact star, and clearly benefits from a large
initial phase space.  One of Case (a) or (b) is also the likely path
for all high--mass X--ray binaries such as Cyg X--1 once the current
wind--fed X--ray phase ends.

Until recently it has generally been assumed that thermal--timescale
episodes are unobservable, as they are short, and without beaming
X--rays could not emerge from the super--Eddington accretion flow at
all. We investigate here the possibility that ULXs could be systems in
this phase, where beaming allows us to see the X--rays.

The thermal--timescale mass transfer rate from a donor near the upper
main sequence is roughly (cf King \& Begelman, 1999)
\begin{equation}
-\dot m_2 \simeq 3\times 10^{-8}m_2^{2.6}\ \msun\ {\rm yr}^{-1}.
\label{tt}
\end{equation}
Comparing with the Eddington accretion rate we can calculate an
acceptance rate 
\begin{equation}
a = 0.43m_1m_2^{-2.6}
\label{a}
\end{equation}
and thus a lifetime (assuming $d=1$)
\begin{equation}
\tau \la 4.3 \times 10^6(b/0.1)^{-1}L_{40}^{-1}m_1m_2^{-1.6}\ {\rm yr}
\label{life}
\end{equation}
and birthrate
\begin{equation}
B \ga 2.3\times 10^{-6}n(b/0.1)L_{40}m_1^{-1}m_2^{1.6}\ {\rm yr}^{-1}
\label{birth}
\end{equation}
per galaxy. In particular, for a system like Cyg X--2, which has $m_1
\simeq 1.4, m_2 \simeq 3\msun$ (King \& Ritter, 1999; Podsiadlowski \&
Rappaport, 2000; Kolb et al., 2000) we get a required birthrate
\begin{equation}
B \ga 1\times 10^{-6}n(b/0.1)L_{40}\ {\rm yr}^{-1}.
\label{cygx2}
\end{equation}
We may compare this with the Galactic birthrate $\sim 10^{-6} -
10^{-7}\ {\rm yr}^{-1}$ deduced for Cyg X--2--like systems (King \&
Ritter, 1999; Podsiadlowski \& Rappaport, 2000; Kolb et al., 2000).
For a high--mass black--hole system like Cyg X--1 both $m_1$ and $m_2$
are probably significantly higher, raising $B$ by as much as an order
of magnitude. However the short X--ray lifetime $\sim 10^5$~yr of this
X--ray phase requires a correspondingly high Galactic birthrate $\sim
{\rm few} \times 10^{-5}~{\rm yr}^{-1}$, again allowing a significant
ULX population. X--ray binaries reach the thermal--timescale phase in
a timescale comparable with the main--sequence lifetime of the
donor. Thus ULXs descending from {\it high--mass} X--ray binaries
would naturally be associated with a young stellar population, as
required by observation.

A possible example of a ULX in the Galaxy is GRS~1915+105, where $L_x
\sim 1\times 10^{39}~{\rm erg\ s}^{-1}$ (e.g. Belloni et al.,
1997). As this is a microquasar, with radio jet axis at about
$70^{\circ}$ to the line of sight (see Mirabel \& Rodriguez, 1999),
only mild beaming $b \sim 0.6$ is possible, even assuming that we view
the system at the edge of the X--ray beam. However this is indeed
sufficient to reduce the luminosity to sub--Eddington values. Moreover
such a geometrical alignment is quite reasonable, as it offers an
explanation for the very unusual long--term behaviour of
GRS~1915+105. The system was not detected in X--rays until 1992, since
when it has remained persistently bright with only short
interruptions. The usual explanation of this as an accretion disc
instability, prolonged by self--irradiation by X--rays (cf King \&
Ritter, 1998) would require an implausibly large disc mass. An
attractive alternative is that the X--ray light curve reflects slight
changes in the X--ray beaming, which would have decreased enough in
1992 to allow us to see the X--rays. 
%Recently, a still better
%candidate for a ULX in the Galaxy has emerged in the shape of another
%microquasar, V4641 Sgr (Orosz et al., 2001). This has a peak X--ray
%luminosity of $\sim 10^{40}$~erg~s$^{-1}$, even closer to the ULXs in
%external galaxies. Encouragingly, the jet axis here is within about
%$6^{\circ}$ of the line of sight, allowing stronger beaming ($b \la
%0.01$). 

\section{CONCLUSIONS}

We have considered models for the ULX class and reached the following
conclusions.

(i) Unbeamed models probably require a black hole of $M_1 \ga
100\msun$ in $\sim 1$~yr binary orbit with an evolved donor
star. Forming such a system presents considerable difficulties, and
even then the likely transient behaviour of the accretion disc in such
a wide system is hard to reconcile with observation. It is still
possible that an {\it individual} ULX may contain a very massive black
hole ($M_1 \ga 3\times 10^4\msun$), perhaps accreting from the
interstellar medium.

(ii) The assumption of mild beaming ($b \sim 0.1 - 0.01$) reduces
$M_1$ to values already observed for Galactic X--ray binaries, and
suggests that ULXs represent a shortlived phase of their
evolution. The most likely candidate for this is the
thermal--timescale mass transfer episode inevitable in a very wide
class of intermediate-- and high--mass X--ray binaries. This in turn
suggests a link to the Galactic microquasars (cf King, 1998, quoted in
Mirabel \& Rodriguez, 1999). The short donor lifetime in high--mass
X--ray binaries would explain why ULXs are associated with young
stellar populations.

The major theoretical uncertainty for (ii) above is whether beaming is
a natural consequence of high accretion rates. Only large numerical
simulations can address this question. Perhaps encouragingly for this
type of model, not only are the X--ray spectra fairly similar to those
of Galactic black hole systems, Kubota et al (2001) observed X--ray
spectral transitions typical of such source in two ULXs.
The same type of spectral and timing variability has also been seen
in the X--9 source in the M81 field (La Parola et al [2001]).
%MBD
There are several possible observational tests of these ideas. First,
continued X--ray monitoring with a view to detecting possible changes
in beaming geometry is clearly worthwhile. We note however that X--ray
eclipses are unlikely in any beamed model, assuming that the X--ray
beam axis is normal to the binary plane. Optical identifications of
ULXs might allow at least two kinds of test: if the total X--ray
luminosities really are as large as predicted if there is no beaming,
one might expect to detect photionization nebulae around ULXs. If on
the other hand ULXs are beamed, and thus of normal stellar mass, one
might hope ultimately to detect a spectroscopic period (say 10's of
days) in a ULX within the Local Group.

\acknowledgements

We thank Mike Garcia, Jim Pringle, Hans Ritter, Tim Roberts, Rashid
Sunyaev and Pete Wheatley for discussions. MBD gratefully acknowledges
the support of a University Research Fellowship from the Royal
Society. Theoretical astrophysics research at Leicester is supported
by a PPARC rolling grant. This work was supported in part by NASA
contract NAS 8-39073 (CXC).


\begin{thebibliography}{}
\bibitem{}
Abramowicz, M.A., Calvani, M., Nobili, L., 1980, ApJ, 242, 772
\bibitem{}
Baraffe,I., Heger, A., Woosley, S.E., 2001, ApJ, in press
(astro--ph/0009458) 
\bibitem{}
Belloni, T., Mendez, M., King, A. R., van der Klis, M.,
van Paradijs, J., 1997, ApJ, 488, 109
\bibitem{} 
Charles, P., 1998, in Theory of Black Hole Accretion Disks,
edited by Marek A.  Abramowicz, Gunnlaugur Bjornsson, and James
E. Pringle. Cambridge University Press, 1998., p.1
\bibitem{}
Dubus, G, Charles, P.A, Long, K., Pasi, J., 1997, ApJ, 490, 47 
%\bibitem{} 
%Dubus, G., Lasota, J.P., Hameury, J.M., Charles, P.A., 1999, MNRAS, 
%303, 139
\bibitem{}
Fabbiano, G., 1989, ARA\&A, 27, 87
\bibitem{}
Fabbiano, G., Zezas, A., Murray, S.S., 2001, ApJ, submitted
\bibitem{}
Frank,J., King, A.R., Raine, D.J., 1992, Accretion Power in
Astrophysics, 2nd Edition, Cambridge University Press, Ch. 4)
%\bibitem{}
%Giles, A.B., Swank, J.H., Jahoda, K., Zhang, W., Strohmayer, T.,
%Stark, M.J., Morgan, E.H., 1996, ApJ, 469, 25
\bibitem{} 
Ho, L., Filippenko, A.V., Sargent, W.L., Peng, C.Y., 1997, ApJS, 112, 391
\bibitem{}
Jaroszynski, M., Abramowicz, M.A., Paczynski, B., 1980, Acta
Astronomica 30, 1
\bibitem{}
Kaaret, P., Prestwich, A.H., Zezas, A., Murray, S.S., Kim, D.W.,
Kilgard, R.E., Schlegel, E.M., Ward, M.J., 2001, MNRAS in press
(astro--ph/0009211)
\bibitem{}
Kalogera, V., Webbink, R.F., 1996, 458, 301
\bibitem{}  
King, A.R., 1988, QJRAS, 29, 1
\bibitem{}
King, A.R., 2000, MNRAS, 315, 306
\bibitem{}
King, A.R., Begelman, M.C., 1999, ApJ, 519, 169
\bibitem{}
King, A.R., Kolb, U., Burderi, L., 1996, ApJ, 464, 127
\bibitem{}
King, A.R., Ritter, H., 1998, MNRAS, 293, L42 
\bibitem{}
King, A.R., Ritter, H., 1999, MNRAS, 309, 253 
\bibitem{}
Kolb, U., Davies, M. B., King, A.R., Ritter, H., 2000, MNRAS 317, 438
\bibitem{}
Kubota, A., Mizuno, T., Makishima, K., Fukazawa, Y., Kotoku, J., 
Ohnishi, T., Tashiro, M., 2001,ApJL, 547L, 119
\bibitem{}
Kulkarni, S. R., Hut P.,  McMillan, S., 1993, Nat., 364, 421
\bibitem{}
La Parola, V., Peres, G., Fabbiano, G., Kim, D.-W., Bocchino, F. 2001, ApJ, in 
press
\bibitem{}
Lee, H.M., 1995, MNRAS, 272, 605
\bibitem{}
Lee, M.H, 1993, ApJ, 418, 147
\bibitem{}
Makishima, Z., Kubota, A., Mizuno, T., Ohnishi, T., Tashiro, M.,
Aruga, Y., Asai, K., Dotani, K., Mitsuda, K., Ueda, Y., Uno, S.,
Yamaoka, K., Ebisawa, K., Kohmura, Y., Okada, K., 2000, ApJ, 535, 632
\bibitem{} 
Matsumoto, H., Tsuru, T. G., Koyama, K., Awaki, H., Canizares, C. R., 
Kawai, N., Matsushita, S., Kawabe, R., 2001, ApJL, 547, L25
\bibitem{}
Matsushita, S., Kawabe, R., Matsumoto, H., Tsuru, T., Kohno, K.,
Morita, K., Okumura, S. K., Vila-Vilaro, B., 2000, ApJ, 545, 107
\bibitem{}
Mirabel, I.F., Rodriguez, L.F., 1999, ARA\&A, 37, 409
\bibitem{}
Orosz, J., et al, 2001, ApJ, in press
\bibitem{}
Papaloizou, J.C.B., 1973, MNRAS, 162, 143
\bibitem{}
Podsiadlowski, Ph., Rappaport, S.A., 2000, ApJ 529, 946
\bibitem{}
Ritter, H., 1999, MNRAS, 309, 360
\bibitem{}
Roberts, T., Warwick, R., 2000, MNRAS, 315, 98 
\bibitem{}
Sigurdsson S., Hernquist, L., 1993, Nat., 364, 423 
\bibitem{}
Stothers, R.B., Chin, C.W., 1999, ApJ, 522, 960
\bibitem{}
White, N.E., Carpenter, G.F., 1978, MNRAS, 183, 11P
\bibitem{}
Zezas, A., Georgantopoulos, I., Ward, M.J., 1999, MNRAS, 308, 302

\end{thebibliography}
\end{document}